\documentclass[twocolumn,prb]{revtex4}%
\usepackage{amsfonts}
\usepackage{amsmath}
\usepackage{graphicx}
\usepackage{amssymb}
\usepackage{dcolumn}
\usepackage{bm}
\usepackage[bookmarks=false]{hyperref}
\usepackage{hypernat}%
\providecommand{\U}[1]{\protect\rule{.1in}{.1in}}
\providecommand{\U}[1]{\protect\rule{.1in}{.1in}}
\providecommand{\U}[1]{\protect\rule{.1in}{.1in}}
\providecommand{\U}[1]{\protect\rule{.1in}{.1in}}
\providecommand{\U}[1]{\protect\rule{.1in}{.1in}}
\providecommand{\U}[1]{\protect\rule{.1in}{.1in}}

\begin{document}
\title{Inherent stochasticity of superconductive-resistive switching in nanowires}
\author{Nayana Shah, David Pekker, Paul M. Goldbart}
\affiliation{Department of Physics, University of Illinois at Urbana-Champaign, 1110 West
Green Street, Urbana, Illinois 61801-3080, USA}
\maketitle

\textbf{Hysteresis in the current-voltage characteristic in a superconducting
nanowire reflects an underlying bistability. As the current is ramped up
repeatedly, the state switches from a superconductive to a resistive one,
doing so at random current values below the equilibrium critical current. Can
a single phase-slip event somewhere along the wire---during which the
order-parameter fluctuates to zero---induce such switching, via the local
heating it causes? We address this and related issues by constructing a
stochastic model for the time-evolution of the temperature in a nanowire whose
ends are maintained at a fixed temperature. The model indicates that although,
in general, several phase-slip events are necessary to induce switching, there
is indeed a temperature- and current-range for which a single event is
sufficient. It also indicates that the statistical distribution of switching
currents initially broadens, as the temperature is reduced. Only at lower
temperatures does this distribution show the narrowing with cooling naively
expected for resistive fluctuations consisting of phase slips that are
thermally activated.}

\begin{figure}[ptb]
\includegraphics[width=8cm]{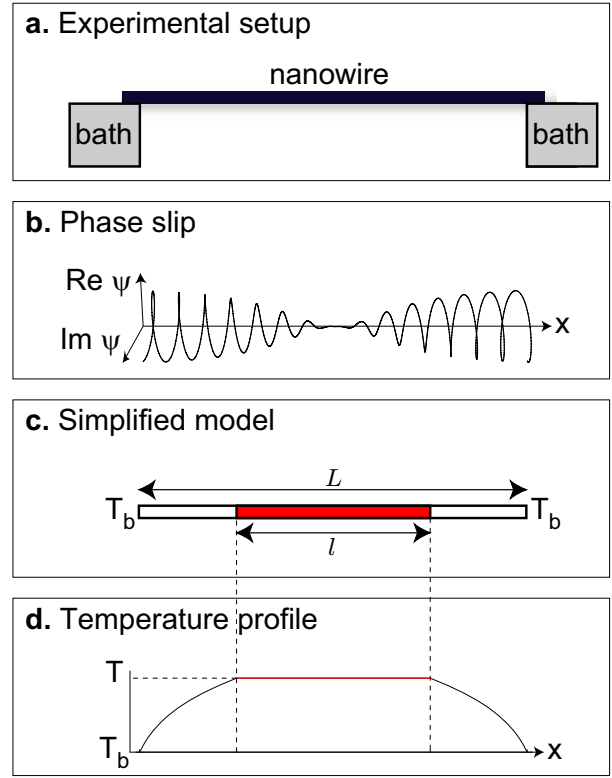}\caption{Model {a.}~Schematic of an
experimental configuration described by our model: a superconducting nanowire
is suspended between two thermal baths. {b.}~Sketch showing the attenuation of
the order parameter in the core of a phase-slip. {c.}~Schematic of the
simplified model. All phase slips are taken to occur in the central
(i.e.~shaded) segment of length $l$, which is assumed to be at a uniform
temperature $T$; heat is carried away through the end segments, which are
assumed to have no heat capacity. The temperature at the ends of the wire is
fixed to be $T_{\text{b}}$. {d.}~Sketch of a typical temperature profile. }%
\label{FIG:schematic}%
\end{figure}

The essential qualitative characteristics of quasi-one-dimensional
superconducting nanowires are controlled by fluctuations of the
superconducting order parameter, these fluctuations being predominantly
thermal or quantal, depending on the temperature regime\cite{SkocpolT1975}.
Bulk superconductors undergo a sharp transition from an electrically
resistanceless (i.e.~superconducting) to a resistive (i.e.~normal) state,
e.g., with increasing temperature. In contrast, as explained by
Little\cite{Little1967} and Langer and Ambegaokar\cite{LangerA1967}, in
quasi-one-dimensional superconductors the resistanceless (and truly
long-range-ordered) state is destabilized by a certain class of accessible
order-parameter fluctuations that connect topologically distinct sectors of
current-carrying states. In so doing, these fluctuations, which are known as
phase-slip events, can dissipate supercurrent, and because of them such
systems undergo a broad evolution between the (nominally) superconductive and
normal states, e.g., with increasing temperature.

Recent advances in the fabrication of ultra-narrow superconducting
wires---using carbon nanotube-\cite{BezryadinLT2000} or
DNA-templating\cite{HopkinsPGB2005}---have spurred renewed interest in
one-dimensional superconductivity and opened up new avenues for investigating
the impact of order-parameter fluctuations. One setting in which
order-parameter fluctuations in superconducting nanowires have been widely
investigated, both theoretically and experimentally, is that of transport
properties in the vicinity of the normal-to-superconducting quantum phase
transition\cite{BezryadinLT2000,LopatinSV2005,RogachevBB2005,ShahL2007,MaestroRSS2007}%
. In this setting, the primary mechanism underlying destruction of (nominal)
superconducting order is depairing associated with magnetic fields or magnetic
impurities. As is well known, applied currents also cause depairing and, if
larger than a certain value (known as the thermodynamic critical or depairing
current), would render the superconducting state locally unstable (regardless
of the role of phase-slip
fluctuations)\cite{Bardeen1962,RomijnKRM1982,Tinkham}.

However, phase-slip fluctuations, which are responsible for the broad
resistive transition in quasi-one-dimensional superconductors, also allow for
premature switching\cite{Giordano1990,TinkhamFLM2003,AltomareCMHT2006} to the
resistive state, i.e.~a nonequilibrium transition from the (nominally)
superconducting, low-resistivity state to the (nominally) normal,
high-resistivity one. If damping of the order-parameter dynamics were low, a
single phase-slip event would induce such switching, in analogy with what
happens in underdamped Josephson junctions. By contrast, nanowires are
generally overdamped, and so, whilst causing resistance, phase slippage does
not, by itself, induce switching. As discussed in Ref.~%
%TCIMACRO{\TeXButton{TinkhamFLM2003}{[\onlinecite{TinkhamFLM2003}]}}%
%BeginExpansion
[\onlinecite{TinkhamFLM2003}]%
%EndExpansion
, this resistance causes Joule heating which, if not overcome sufficiently
rapidly by conductive cooling, effectively reduces the depairing current,
ultimately to below the applied current, thus causing switching to the highly
resistive state. Naturally, this switching is not deterministic, owing to the
underlying stochasticity of the phase-slip events that are responsible for the
resistance. Rather, for a given subcritical applied current there is a
statistical distribution of times at which switching occurs, characterized by
a mean switching time (i.e.~a superconducting state \textquotedblleft
lifetime\textquotedblright).

Our focus here is on \textit{stochastic\/} aspects of the
superconducting-to-resistive switching dynamics, an area that has not received
much attention, to date. \textit{Inter alia\/}, by obtaining the
current-dependent mean switching time and convolving it with the sweep rate of
the applied current that describes the experimental protocol, we shall
determine the statistical distribution of currents at which switching occurs.
Besides its fundamental significance, the characterization of switching
dynamics in nanowires seems likely to have technological implications, such as
for the integration of superconducting wires into electronic circuitry as
controllable (current-limiting) switching elements, the implementation of
nanowire-based devices\cite{HopkinsPGB2005,PekkerBHG2005,JohanssonSSJT2005},
and the exploration of the use of nanowires in quantum computers.

\begin{figure}[ptb]
\includegraphics[width=8cm]{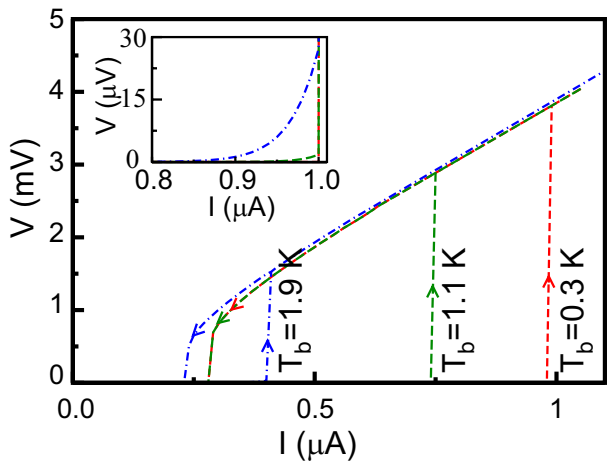}\caption{{ Hysteresis in the I-V characteristic obtained by finding
the steady-state solutions of Eq.~(\ref{HCE}) in the continuous Joule heating
limit. }}%
\label{FIG:hysteresis}%
\end{figure}

Having in mind the configuration in recent and ongoing experiments on
superconducting nanowires, we consider a free-standing wire of length $L$ and
cross-sectional area $A$, the ends of which are held at a fixed temperature
$T_{\text{b}}$, as shown in Fig.~\ref{FIG:schematic}. The fact that the wire
is free-standing (i.e.~lacks any substrate) is conducive to a clear
interpretation of the measurements. On the other hand, the absence of an
overall thermal bath means that any heat generated locally in the wire by a
source term $Q\ $can be taken away only through the ends; the corresponding
heat conduction equation for the temperature ${\Theta}(x,t)$ at position $x$
along the wire at time $t$ reads%
\begin{equation}
C_{\text{v}}({\Theta})\,\partial_{t}{\Theta}(x,t)=\partial_{x}\left[
K_{\text{s}}({\Theta})\,\partial_{x}{\Theta}(x,t)\right]  +Q(x,t), \label{HCE}%
\end{equation}
and is characterized by the specific heat $C_{\text{v}}({\Theta})$ and thermal
conductivity $K_{\text{s}}({\Theta})$ of the wire, together with the boundary
condition ${\Theta}(\pm L/2,t)=T_{\text{b}}$ at its ends. Note that although
our analysis rests on the premise that there are no additional heat-removing
channels, it can readily be extended to account for such possibilities.

Before addressing dynamical issues, let us dwell briefly on the steady-state
solutions of the heat conduction equation, obtained by setting $\partial
_{t}{\Theta}(x,t)=0$ and assuming that the wire is subjected to temporally
continuous Joule heating at a rate given by $ALQ(x)=I^{2}R({\Theta}(x),I)$.
Here, the function $R({\Theta}^{\prime},I)$ is to be understood as the
resistance of an entire wire held at a uniform temperature ${\Theta
}(x)={\Theta}^{\prime}$. The system-wide I-V characteristic at a given
boundary value $T_{\text{b}}$ can be traced by obtaining the temperature
profile ${\Theta}(x)\ $for every value of current $I$ in both up and down
(parametric)\ sweeps of $I$. On determining $R({\Theta},I)$ via the
current-biased version of LAMH
theory\cite{LangerA1967,McCumber1968,McCumberH1970} for phase-slips, the
$I$-$V$ curves are indeed found to become progressively more hysteretic in
$I\ $as $T_{\text{b}}$ is lowered (see Fig.~\ref{FIG:hysteresis}). This
steady-state problem was previously studied by Tinkham et
al.\cite{TinkhamFLM2003}, who used for $R({\Theta},I)$ the experimental
linear-response resistance measured at $T_{\text{b}}={\Theta}$, leading to
qualitative agreement with the hysteresis observed in MoGe
nanowires\cite{TinkhamFLM2003}.

Our aim here is to study the inherent stochasticity in the switching process,
and therefore it is necessary for us to explicitly take into consideration the
fact that the resistive fluctuations of the superconducting nanowire consist
of discrete phase-slip events (labelled by $i$) that take place at random
moments of time $t_{i}$ and are centered at random spatial locations $x_{i}$.
The work done on the wire by a phase slip may be obtained from the time
integral of $IV(t)$, in which the Josephson relation $d\phi/dt=$ $2eV/\hslash$
may be used to relate the voltage pulse to the rate of change of the phase
difference\cite{Tinkham}, via fundamental constants $\hslash$ and $e$. Hence,
a single phase slip (or anti-phase slip), which corresponds to a decrease (or
increase) of $\phi$ by $2\pi$, will heat (or cool) the wire by a
\textquotedblleft quantum\textquotedblright\ of energy $hI/2e$. Thus we arrive
at the central thrust of our paper: the dynamics of switching from the
superconducting to the resistive state in the nanowires is controlled by a
heat conduction equation that is stochastic by virtue of its source term:
\begin{equation}
Q(x,t)=\frac{hI}{2e}\frac{1}{A}\sum_{i}\sigma_{i}F(x-x_{i})\delta(t-t_{i}),
\end{equation}
where $F(x-x_{i})$ is a normalized (to unity) form factor representing the
relative spatial distribution of heat produced by the $i^{\text{th}}$
phase-slip event, and $\sigma_{i}=\pm1$ for phase (anti-phase) slips. The
probability per unit time $\Gamma_{\pm}$ for an anti-phase (phase) slip to
take place, depends on the local temperature ${\Theta}(x,t)$ and the current
$I$. The randomness in $x_{i}$ and $t_{i}$ generates the stochasticity in the
switching from the superconductive to the resistive state.

To capture the essential physics whilst making the problem amenable to
analysis, we shall consider the simpler model, represented in
Figs.~\ref{FIG:schematic}c and \ref{FIG:schematic}d. Given that the edge
effects favor phase-slip locations away from the wire ends, the source term is
restricted to the region near the center of the wire. The system is thus
modeled by assuming that (i) the heating takes place within a central segment
of length $l$ to which a uniform temperature $T$ is assigned, and (ii)\ the
heat is conducted away through the end segments, within which we ignore the
heat capacity\cite{fn1}. To simplify the problem further, we make use of the
fact that $\Gamma_{+}\ll\Gamma_{-}$ and ignore the process of cooling by
anti-phase slips. To account indirectly for their presence, we use a reduced
rate $\Gamma\equiv\Gamma_{-}-\Gamma_{+}$ instead of $\Gamma_{-}$ for
phase-slip events. This ensures that the discrete expression for $Q$ will
correctly reduce to the continuous Joule-heating expression, in view of the
LAMH formula $R(T,I)=$ $h\Gamma/2eI$.

By using the model defined above, the description reduces to a stochastic
ordinary differential equation for the time-evolution of the temperature of
the central segment:
\begin{equation}
\frac{dT}{dt}=-\alpha(T,T_{\text{b}})(T-T_{\text{b}})+\eta(T,I)\sum_{i}%
\delta(t-t_{i}), \label{SDE}%
\end{equation}
where the second term on the RHS corresponds to heating by phase slips, and
the first term to cooling as a result of conduction of heat from the central
segment to the external bath via the two end-segments, each of length
$(L-l)/2$. The temperature-dependent cooling rate $\alpha$ is given by
\begin{equation}
\alpha(T,T_{\text{b}})\equiv\frac{4}{l(L-l)C_{\text{v}}(T)}\frac
{1}{T-T_{\text{b}}}\int_{T_{\text{b}}}^{T}dT^{\prime}\,K_{\text{s}}(T^{\prime
})\text{.}%
\end{equation}
If $T_{\text{i}}$ and $T_{\text{f}}$ are temperatures before and after a phase
slip then, using
\begin{equation}
A\text{ }l\int_{T_{\text{i}}}^{T_{\text{f}}}C_{\text{v}}(T^{\prime
})\,dT^{\prime}=\frac{hI}{2e},
\end{equation}
we can express the temperature `impulse' due to a phase slip,i.e$~T_{\text{f}%
}-T_{\text{i}}\equiv\eta(T_{\text{i}},I)\equiv\widetilde{\eta}(T_{\text{f}%
},I)$, as function of either $T_{\text{i}}$ or $T_{\text{f}}$, depending on
the context.

\begin{figure}[ptb]
\includegraphics[width=8cm]{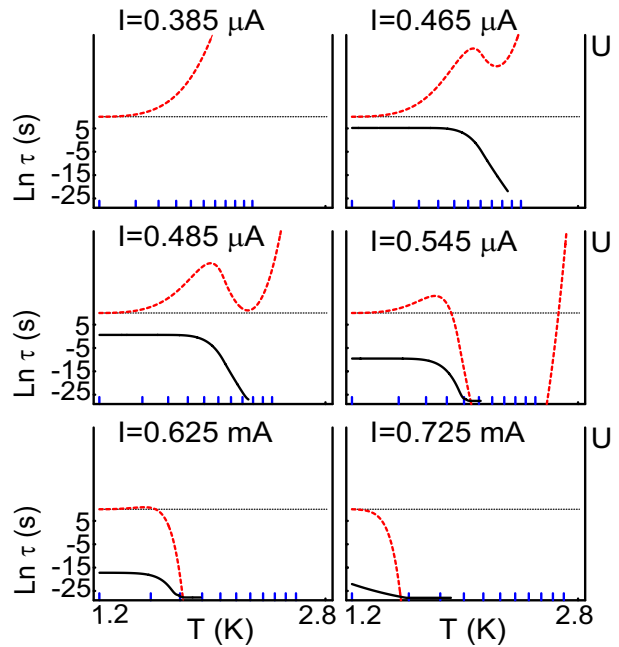}\caption{Effective potential
$U(T,T_{b},I)$ (dashed line) and the mean first-passage time $\tau$ as
functions of the temperature $T\ $of the central segment for various bias
currents $I$ and for $T_{\text{b}}=1.2\,$K. The marks on the temperature axis
indicate the temperatures that the central segment would have after
$1,2,...,10$ phase slips in the absence of cooling (as given by
Eq.~(\ref{sequence}) for $T_{0}=T_{\text{b}}$).}%
\label{FIG:potential}%
\end{figure}

Let us now elucidate the physical and mathematical structure of Eq.~(\ref{SDE}%
). To begin with, we shall consider the continuous-heating limit,
$\eta(T,I)\Gamma(T,I)$, for the source term, and express Eq.~(\ref{SDE}) as
$dT/dt=-\partial U/dT$. In Fig.~\ref{FIG:potential}, we illustrate the form of
the `potential' $U(T,T_{\text{b}},I)$ for fixed $T_{\text{b}}$: there is a
range of currents $I$ for which $U$ has two local minima, corresponding to the
superconducting (at low-$T$) and the resistive (at high-$T$) states, separated
by a local maximum. The resulting bistability is central to the underlying
physics. On the one hand, it explains the origin of the hysteretic behavior;
on the other hand, it provides a basis for phrasing the question of stochastic
switching dynamics in superconducting nanowires in terms of an existing
general framework for stochastic bistable systems. In what follows, we focus
on the stochastic variable $T(t)$; to ease the notation we do not display the
dependences on $I$ and $T_{\text{b}}$ unless essential. To continue the
analysis of the stochastic equation, imagine turning off the cooling term. If
we now start with an initial temperature $T_{0}$ then
\begin{equation}
T_{0},T_{0}+\eta(T_{0}),T_{0}+\eta(T_{0})+\eta(T_{0}+\eta(T_{0})),...
\label{sequence}%
\end{equation}
defines the discrete sequence of values that $T$ jumps to, as marked on the
horizontal axes in Fig.~\ref{FIG:potential} for $T_{0}=T_{\text{b}}$. The
probability per unit time, $\Gamma(T)$, to make a jump changes at each step,
and so does the size $\eta(T)$ of the jump, owing to their explicit dependence
on temperature. On the other hand, if we turn off the heating term then we
have a deterministic problem in which $T\ $would decay at a rate $\alpha(T)$,
from its initial value $T_{0}>T_{\text{b}}$ to the bath temperature
$T_{\text{b}}$, which is the lowest value $T\ $can have. It is the competition
between the discrete heating and the continuous cooling that makes for a
rather rich stochastic problem. We hope that our solution will also furnish
insight into other physical problems that possess a similar mathematical
structure. 

The master equation for $P(T,t)$, the probability for the temperature of the
central segment of the nanowire to be $T$ at a time $t$ (given that it had
some initial value $T_{0}$ at time $t_{0}$), reads
\begin{align}
\partial_{t}P(T,t)  &  =\partial_{T}\,[(T-T_{\text{b}})\,\alpha
(T)\,P(T,t)]-\Gamma(T)\,P(T,t)\nonumber\\
&  +\Gamma\big (T-\widetilde{\eta}(T)\big)\,P\big (T-\widetilde{\eta
}(T),t\big)\,\big (1-\partial_{T}\,\widetilde{\eta}(T)\big)\text{,} \label{ME}%
\end{align}
where the first (i.e.~the transport) term corresponds to the effect of
cooling, and the last two terms correspond to the effects of heating. Note
that the term $(1-\partial_{T}\widetilde{\eta}(T))$ appears because of the
dependence of the jump size on $T$, as given by $\widetilde{\eta}(T)$. The
fundamental quantity of interest is the mean switching time $\tau_{\text{s}%
}(T_{\text{b}},I),$ i.e.~the mean time required for the wire to switch from
being superconductive to resistive, assuming that the wire has temperature
$T=T_{\text{b}}$ when the current $I$ is turned on at time $t=0$. The master
equation, Eq.~(\ref{ME}), provides the starting point for generalizing the
standard procedure for computing $\tau_{\text{s}}$ via the evaluation of the
mean first-passage time\cite{Kampen}.

The mean first-passage time $\tau(T\rightarrow T^{\text{*}})$, to go past a
point $T=T^{\text{*}}$ for the first time having started from some
$T<T^{\text{*}}$, can be shown to satisfy the equation%
\begin{equation}
-(T_{\text{b}}-T)\,\alpha(T)\,\partial_{T}\,\tau(T)+\Gamma(T)\left[
\tau\big (T\big)-\tau\big (T+\eta(T)\big)\right]  =1, \label{MFPT}%
\end{equation}
together with the conditions $\tau(T)=0$ for $T>T^{\text{*}}$ and
$d\tau(T)/dT=0$ at $T=T_{\text{b}}$, which are appropriate for our problem.
Some illustrative plots for $\tau(T\rightarrow T^{\text{*}})$, obtained by
numerically solving Eq.~(\ref{MFPT}) are shown in Fig.~\ref{FIG:potential},
with the choice of $T^{\text{*}}$ being somewhat larger than the location of
the local maximum of $U$. From these plots we see that the mean first-passage
time has a plateau at low values of $T\ $and then rapidly decreases in the
vicinity of the potential barrier. At very high currents, as can be seen from
the last panel of Fig.~\ref{FIG:potential}, the local stability of the
superconducting state disappears, and so does the plateau in the mean
first-passage time. In these plots, the tick marks on the $T$ axes correspond
to the temperatures given by the sequence~(\ref{sequence}) for $T_{0}%
=T_{\text{b}}$.

As long as the high-$T$ minimum is lower than the low-$T$ one, and
$T^{\text{*}}$ is chosen to be appreciably past the intervening potential
maximum (in order to eliminate the possibility of reversion to the
superconducting state), we can make the identification $\tau_{\text{s}%
}(T_{\text{b}},I)\equiv\tau(T_{\text{b}}\rightarrow T^{\text{*}},T_{\text{b}%
},I)$. The number of tick marks (see sequence~(\ref{sequence})) between
$T_{\text{b}}$ and $T^{\text{*}}$ is nothing but the of number $N(T_{\text{b}%
},I)$ of phase-slip events required to raise the temperature of the central
segment from $T_{\text{b}}$ to $T^{\text{*}}$ in the absence of cooling.
Accordingly, $N(T_{\text{b}},I)$ also provides an estimate of the number of
phase-slip events needed to overcome the potential barrier if the timespan of
these events is insufficient to allow significant cooling to occur. `Thermal
runaway'---heating by rare sequences of closely-spaced phase slips that
overcome the potential barrier---constitutes the mechanism of
superconductive-to-resistive switching within our model. As the $N(T_{\text{b}%
},I)$ becomes large, the total number of phase-slip events taking place before
switching can happen, and correspondingly the value of $\tau_{\text{s}%
}(T_{\text{b}},I)$, may indeed be quite large. \begin{figure}[ptb]
\includegraphics[width=8cm]{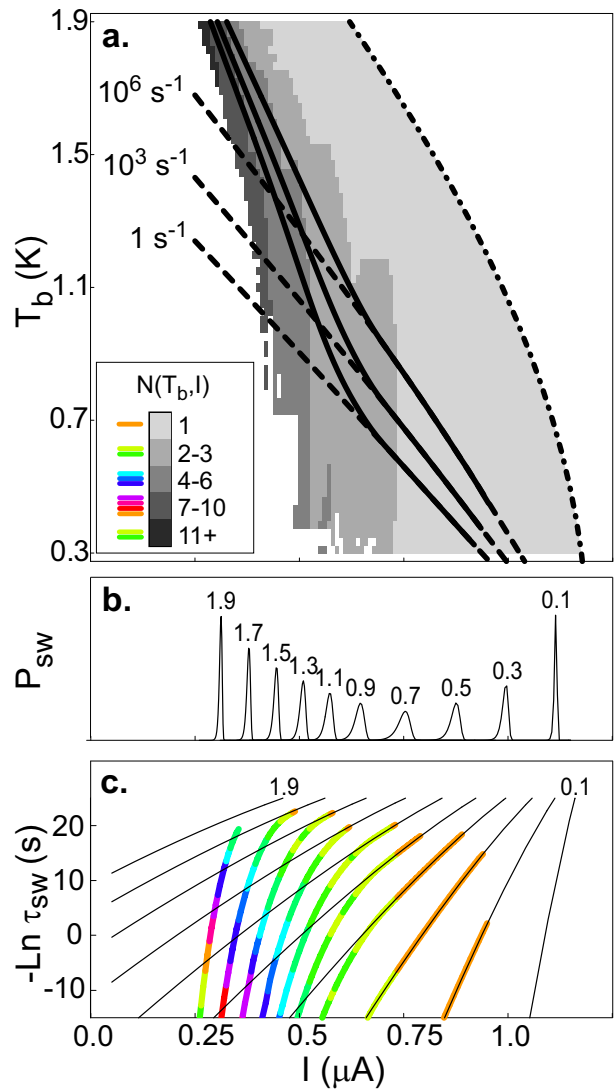}\caption{Switching statistics {a.~}Map showing
$N(T_{\text{b}},I)$ (see text) and the contour lines (solid lines) for the
inverse of mean switching time, $\tau_{\text{s}}^{-1}=1,$ $10^{3},$
$10^{6}\text{ s}^{-1}$; the contour lines (dashed lines)\ for the phase-slip
rate $\Gamma$ are also shown. The thermodynamic (depairing) critical current
(dashed-dotted line) is plotted for reference. {b.}~Switching-current
distributions $P_{\text{SW}}$ obtained at various values of $T_{\text{b}}$ and
for $r=58\,\mu\text{A}/\text{s}$. {c.~}The logarithms of $\tau_{\text{s}}%
^{-1}$ (colored lines) and of $\Gamma$ (thinner black lines) as a function of
$I$, obtained for the same set of $T_{\text{b}}$values as in panel~(a). The
colors of $\tau_{\text{s}}^{-1}$ plots correspond to different values of
$N(T_{\text{b}},I)$ [as indicated in the legend of panel~(a)].}%
\label{FIG:mean-switching-time}%
\end{figure}

Our key findings are summarized in Fig.~\ref{FIG:mean-switching-time}. There
is a region of $I\ $and $T_{\text{b}}$ for which the occurrence of just one
phase slip is sufficient to cause the nanowire to switch from the
superconductive to the resistive state\cite{fn2}; in this case $\tau
_{\text{s}}^{-1}=\Gamma$. A switching measurement in this range can thus
provide a way of detecting and probing a single phase-slip fluctuation. As,
outside this range, several phase-slip events are required for switching,
$\tau_{\text{s}}^{-1}$ deviates from $\Gamma$ (see
panel~\ref{FIG:mean-switching-time}c). A graphical representation of the
contour lines for a few values of $\tau_{\text{s}}^{-1}$ and $\Gamma$, chosen
in an experimentally accessible range, is provided in
panel~\ref{FIG:mean-switching-time}a. Whilst the spacing between the $\Gamma$
contour lines decreases monotonically on lowering $T_{\text{b}}$, the spacing
between $\tau_{\text{s}}^{-1}$ lines can be seen to behave non-monotonically.

The mean switching time $\tau_{\text{s}}$ in bistable current-biased systems
can be either directly measured or extracted from the switching-current
statistics\cite{FultonD1974} generated via the repeated tracing of the
$I$-$V\ $characteristic by ramping the current up and down at some sweep rate
$r$. For this reason, in Fig.~\ref{FIG:mean-switching-time}b\ we have
illustrated the behavior of this distribution of switching currents in
superconducting nanowires based on the theory presented here. Upon raising
$T_{\text{b}}$, one would naively expect the distribution to become broader
for a model involving thermally activated phase slips. Such an broadening in
the distribution-width is indeed obtained up to a crossover temperature scale
$T_{\text{b}}^{\text{cr}}(r)$ (i.e.~the temperature below which, loosely
speaking, switching is induced by single phase slips). However, on continuing
to raise $T_{\text{b}}$, but now through temperatures above $T_{\text{b}%
}^{\text{cr}}(r)$, the distribution-width shows a seemingly anomalous
decrease. This is a manifestation of the now-decreasing spacing between the
$\tau_{\text{s}}$ contour lines. This striking behavior above $T_{\text{b}%
}^{\text{cr}}$ may be understood by the following reasoning: the larger the
number of phase-slips in the sequence inducing the
superconductive-to-resistive thermal runaway, the smaller the stochasticity in
the switching process and, hence, the sharper the distribution of switching
currents. This non-monotonicity in the temperature dependence of the width of
switching-current distribution, along with the existence of a regime in which
a single phase-slip event can be probed, are the two key predictions of our theory.

We gratefully acknowledge invaluable discussions with A. Bezryadin, M. Sahu, and T-C. Wei, and, on the issue of numerical approaches, with B. K. Clark.  This work was supported by the DOE Division of Materials Sciences under Award No.~DE-FG02-07ER46453, through the Frederick Seitz Materials Research
Laboratory at the University of Illinois at Urbana-Champaign, and by NSF grant DMR 0605813.


\begin{thebibliography}{99}
\bibitem {fn1}{The length $l$ is larger than coherence length and can be
roughly estimated to be of the order of charge imbalance length required for
thermalization. Note that the effective length $L$ within the model may differ
from the geometric length of the wire.}
\bibitem {fn2}{Note that within the regime described by our model, i.e. for
$r^{-1}$ greater than the Ginzburg Landau time, the depairing critical current
is never reached, as we expect a single-phase to have already triggered a
switching event.}
\end{thebibliography}
\end{document}